# A conservative diffuse interface method for two-phase flows with provable boundedness properties

Shahab Mirjalili[a], Christopher B. Ivey[b], Ali Mani[a,*]

[a]*Department of Mechanical Engineering, Stanford University, Stanford, CA 94305, USA*
[b]*Cascade Technologies, Inc., Palo Alto, CA 94303, USA*

## Abstract

Central finite difference schemes have long been avoided in the context of two-phase flows for the advection of the phase indicator function due to numerical overshoots and undershoots associated with their dispersion errors. We will show however, for an incompressible flow, in the context of a specific diffuse interface model, one can maintain the boundedness of the phase field while also taking advantage of the low cost and ease of implementation of central differences to construct a non-dissipative discretization scheme for the advective terms. This is made possible by combining the advection and reinitialization steps of a conservative level set scheme introduced by Olsson and Kreiss [J. Comput. Phys., 210, 225 (2005)] to form a phase field equation similar to that of Chiu and Lin [J. Comput. Phys., 230, 185 (2011)]. Instead of resorting to specialized upwind methods as in these articles, we prove that the boundedness of the phase field is guaranteed for certain choices of the free parameters ($\epsilon$ and $\gamma$) for a specific central difference scheme that we propose. The proposed discretely conservative and bounded phase field equation, which is free of any reinitialization or mass redistribution, possesses desirable properties that can be leveraged in the coupled finite difference discretization of the two-phase momentum equation. Additionally, as compared to the state-of-the-art conservative and bounded two-phase flow methods, the proposed method boasts competitive accuracy-vs-cost trade-off, small memory requirements, ease of implementation and excellent parallelizability, providing a viable alternative for realistic two-phase flow calculations.

## 1. Introduction

Conservative level set methods have gained popularity since Olsson and Kreiss [1] represented the sharp interface between two immiscible phases with an artificially smoothened profile that takes the form of $\phi(\vec{x}) = 0.5(1 + \tanh(s(\vec{x})/2\epsilon))$, where $\vec{x}$ is the position, $s(\vec{x})$ is the signed distance from the interface and $\epsilon$ is a parameter controlling the thickness of the smoothed interface. Since then, the conservative level set method and its proposed reinitialization equation have been improved in various works [2, 3, 4]. All of these works took advantage of diffusive and sharpening (compressive) terms in an iterative reinitialization procedure to ensure convergence to the aforementioned $\phi(\vec{x})$ profile. The hyperbolic tangent shape of the interface in the conservative level set method resembles the thermodynamically derived equilibrium profile in phase field models [5, 6, 7, 8]. Most of the older phase field models were either based on the Cahn-Hilliard or the Allen-Cahn equation. Among these, the Cahn-Hilliard model has gained popularity in the two-phase flow community due to its conservative properties and other beneficial traits stemming from a defined energy norm [9, 10]. However, dealing with the fourth-order derivative on its right-hand side (RHS)

---

[*]Corresponding author: Ali Mani
  *Email addresses:* `ssmirjal@stanford.edu` (Shahab Mirjalili), `civey@cascadetechnologies.com` (Christopher B. Ivey), `alimani@stanford.edu` (Ali Mani)

renders the Cahn-Hilliard equation challenging for simulations of realistic two-phase flows. Even in the absence of discretization challenges, at the PDE level the Cahn-Hilliard model fails to guarantee bounded solutions, potentially resulting in negative densities [11, 12]. As an alternative approach and inspired by the conservative level set of Olsson and Kreiss [1], Chiu and Lin [13] obtained a conservative phase field model that only requires second derivatives and asymptotes exactly to $\phi = 1$ or $\phi = 0$ far from the interface. For an incompressible velocity field, their model can be written as

$$\frac{\partial \phi}{\partial t} + \nabla \cdot (\vec{u}\phi) = \nabla \cdot \left[ \gamma \left( \epsilon \nabla \phi - \phi(1-\phi) \frac{\nabla \phi}{|\nabla \phi|} \right) \right]. \tag{1}$$

Chiu and Lin [13] opted to use a dispersion-relation-preserving dual-compact upwind scheme for advection of the phase field on a staggered mesh and fixed their RHS parameters at $\epsilon = 0.7\Delta x$ and $\gamma = u_{max}$. With such choices, inspite of their tailored advection operators, their discretization led to overshoots and undershoots. Overshoots and undershoots necessitate clipping or mass redistribution algorithms, both of which are unphysical treatments whose impact is amplified in large density ratio flows.

For single-phase incompressible flows, second-order central difference advective operators on staggered grids have been shown to naturally provide discrete conservation of mass, momentum and energy, provided that continuity is satisfied exactly [14]. At first sight, this may hint at the potential benefits of using second-order central differences for the advection of the phase field; however, central differences have been traditionally excluded from discretizations of the phase advection since they can excite overshoots and undershoots, particularly near sharp boundaries due to their dispersion error.

The prospect of a discretely conservative, bounded advection scheme prompted us to revisit Equation 1 and examine how the choice of $\epsilon$ and $\gamma$ can impact the boundedness of $\phi$. Interestingly, we find that central difference schemes when used for Equation 1, as opposed to the advection equation, provide provable boundedness properties.

We note that applying central differences to advective fluxes in the presence of artificial diffusivity (first term on the right-hand side of Equation 1) is reminiscent of upwind schemes and/or artificial viscosity schemes adopted in compressible flows. However, unlike such numerical treatments, Equation 1 involves an additional nonlinear artificial term (second term on the right-hand side of Equation 1), which opposes the artificial diffusion. Therefore, it is highly crucial to discretely analyze the impact of the artificial diffusivity coupled with the nonlinear term in Equation 1. In fact, our analysis reveals a non-trivial crossover line in the two-dimensional $\epsilon - \gamma$ parameter space, above which, stability requirements on the time step, $\Delta t$, from Equation 1 lead to boundedness of $\phi$ for an incompressible flow.

In what follows, we first present proof of boundedness in detail for one-dimensional (1D) constant velocity systems and then extend the proof to multi-dimensional flows with arbitrary incompressible velocity fields. We also initially prove boundedness when Equation 1 is integrated in time using the explicit Euler scheme; subsequently, we will show the resulting boundedness criteria maintains boundedness for higher-order explicit Runge-Kutta (RK) time-stepping schemes as well. We confirm our analytical predictions with numerical studies that use various time-integration schemes in 1D and 2D settings. Numerical tests are also included to demonstrate the convergence properties and accuracy of the proposed discretization. Finally, we conclude by summarizing our findings and highlighting the desirable attributes of the proposed two-phase flow simulation method.

## 2. Analysis

In order to analytically investigate the boundedness properties of the numerical solution to Equation 1 when central differences are used for spatial derivatives, we initially make some simplifying assumptions on the flow and time-integration scheme. Namely, we study the problem of 1D advection with constant velocity using explicit Euler time-stepping. We readily extend our proof to multi-dimensional non-uniform incompressible flows and then provide arguments as to why the boundedness criteria found using explicit Euler are conservative, allowing for our findings to be applied to higher order time-stepping schemes.

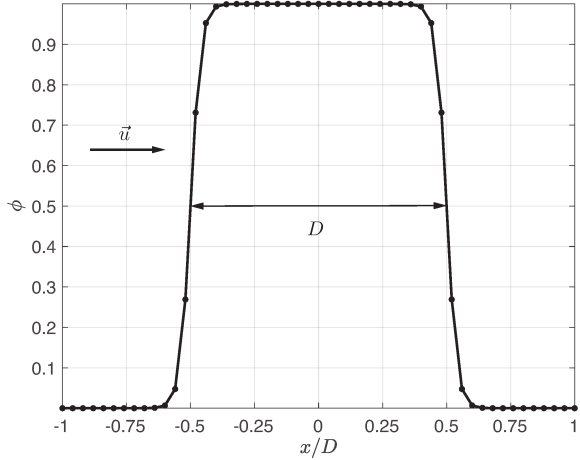

Figure 1: Initial $\phi$ field for the problem of a 1D drop of diameter $D$ being advected with a constant positive velocity $u$ in a periodic domain of length $2D$.

In the phase field equation presented in 1, the parameters $\epsilon$ and $\gamma$ have positive constant values. Since we only concern ourselves with incompressible flows, any analysis in 1D assumes constant velocity, $\vec{u}$, which for convenience we assume to be positive. To confirm our analytical findings, we also present numerical results by performing simulations of a drop with diameter $D$, advected with a constant positive velocity, $u$, on a uniform mesh, $\Delta x$, in a periodic domain of length $2D$ (see Figure 1), while $\epsilon$ and $\gamma$ are fixed parameters. The 1D drop is initialized with the appropriate tangent hyperbolic interface shapes on its left and right interface.

In Section 2.1, we will present requirements for numerical stability of Equation 1 and motivate boundedness analysis of $\phi$. Then in Section 2.2, the boundedness requirements for Equation 1 are presented and proven. Through out this section, plots of the companion numerical simulations are provided as corroboration for our analytic studies.

2.1. Stability analysis

By recognizing the advective nature of the sharpening term, for an incompressible flow, one can rewrite Equation 1 in the following form,

$$\frac{\partial \phi}{\partial t} + (\vec{u} + \gamma \hat{n}(1 - 2\phi)) \cdot \nabla \phi = \gamma \epsilon \nabla^2 \phi, \qquad (2)$$

where $\hat{n} = \nabla \phi / |\nabla \phi|$. We note that Equations 1 and 2 are not equivalent in discrete form, but only analytically and in regions where $\hat{n}$ does not change sign. In this section, we use Equation 2 to only develop physical insights given its appropriate form for analysis using method of characteristics. Equation 2 shows that in regions of the domain where $\vec{u}$ and $\gamma \hat{n}(1 - 2\phi)$ are of the same sign, the characteristics can move faster than the flow itself, suggesting a modified time step requirement for stability, $\Delta t_{mod}(u, \Delta x, \epsilon, \gamma)$. The characteristic's speed from Equation 2 clearly depends on the bounds of $\phi$. Since $\phi$ is bounded between 0 and 1 initially, and should ideally remain in those bounds for two-phase simulations, we will continue our stability analysis with that assumption, yielding $\max(|1 - 2\phi|) \leq 1$. Therefore, from temporal stability analysis point of view, the maximum characters speed is $u + \gamma$. As explained before, without loss of generality, we focus on Euler time integration. We will use the modified wavenumber analysis to study the stability of applying Euler time stepping and second-order central differences to Equation 2. Accordingly, we semi-discretize Equation 2 and substitute $\phi_j(t) = \rho(t)e^{ikx_j}$ to study the growth of $\rho(t)$ at different wave numbers. If we denote the number of mesh points in the domain to be $N$ and $n \in \{0, 1, 2, ..., N/2\}$, then for wave numbers given by $k = 2\pi n/(N\Delta x)$:

$$\frac{d\rho}{dt} = [-(u+\gamma)\frac{\sin(k\Delta x)}{\Delta x}i - \frac{2\gamma\varepsilon}{\Delta x^2}(1-\cos(k\Delta x))]\rho = \lambda\rho \qquad (3)$$

For Euler time stepping, $1 + \lambda\Delta t = \sigma$ is the amplification factor, and stable numerical time integration of Equation 2 requires

$$\left|-(u+\gamma)\frac{\sin(k\Delta x)}{\Delta x}\Delta t i + 1 - \frac{2\gamma\varepsilon}{\Delta x^2}(1-\cos(k\Delta x))\Delta t\right| = |\sigma(u, \Delta x, \epsilon, \gamma, k)| \leq 1. \qquad (4)$$

Equation 4 states that after one chooses the values of $\epsilon$ and $\gamma$, $\Delta t$ should be chosen such that for a specified advection velocity ($u$) on a uniform mesh ($\Delta x$), the amplification factor at all above wave numbers would be equal to or less than one. If we define the non-dimensional parameters $\alpha = (u+\gamma)\Delta t/\Delta x$ and $\beta = 2\gamma\epsilon/[(u+\gamma)\Delta x]$, then

$$|-\alpha\sin(k\Delta x)i + 1 - \alpha\beta(1-\cos(k\Delta x))| \leq 1, \qquad (5)$$

which with some algebra can be shown to be equivalent to

$$\alpha \leq \frac{\beta}{\cos(k\Delta x/2)^2 + \beta^2\sin(k\Delta x/2)^2}, \qquad (6)$$

for $k = 2\pi n/(N\Delta x)$, with $n \in \{0, 1, 2, ..., N/2\}$. After choosing $\epsilon$ and $\gamma$ for a mesh, the minimum value of the right hand side term in Equation 6 for different wave-numbers determines the supremum of $\alpha$, which gives the time-step requirement for stability. It is easy to verify that

$$\alpha \leq min(\frac{1}{\beta}, \beta). \qquad (7)$$

Using the definitions of $\alpha$ and $\beta$ we obtain the time-step requirement for stability:

$$\Delta t \leq \Delta t_{mod} \equiv min\left(\frac{\Delta x^2}{2\gamma\epsilon}, \frac{2\gamma\epsilon}{(u+\gamma)^2}\right). \qquad (8)$$

Equation 7 also reveals that the maximum stable time step, for fixed $\gamma$, corresponds to $\alpha = \beta = 1$. Specifically, $\beta = 1$ when

$$\epsilon/\Delta x = \frac{\gamma/u + 1}{2\gamma/u}, \qquad (9)$$

leading to the time-step requirement of

$$\Delta t \leq \Delta t_{mod,max} = \frac{\Delta x}{u+\gamma}. \qquad (10)$$

Equation 9 defines a line in the dimensionless $\epsilon/\Delta x - \gamma/u$ parameter space, shown in Figure 2a, where the discretized partial differential equation (PDE) in Equation 1 changes nature. For parameters above this line, the diffusion term overcomes the advective and sharpening (compressive) terms in determining numerical stiffness. Below this line, Equation 1 is more hyperbolic than parabolic as the physical advection and sharpening terms dominate in controlling numerical stiffness. As shown in Equation 10, for a given $\gamma/u$, choosing $\epsilon/\Delta x$ to lie on this line is beneficial as it allows for maximum stable time-step. We refer to this line as the crossover line, and we will encounter it again as it is central to our boundedness analysis.

The derived temporal stability criterion (Equation 8) was obtained for numerical solution to Equation 2, which is discretely different from the original phase field PDE, Equation 1. However, our numerical investigations over a wide range of non-dimensional parameters, $\epsilon/\Delta x$ and $\gamma/u$, confirm that this stability criterion holds for Equation 1 as well. For the rest of this work, we focus our attention on the numerical solution of Equation 1.

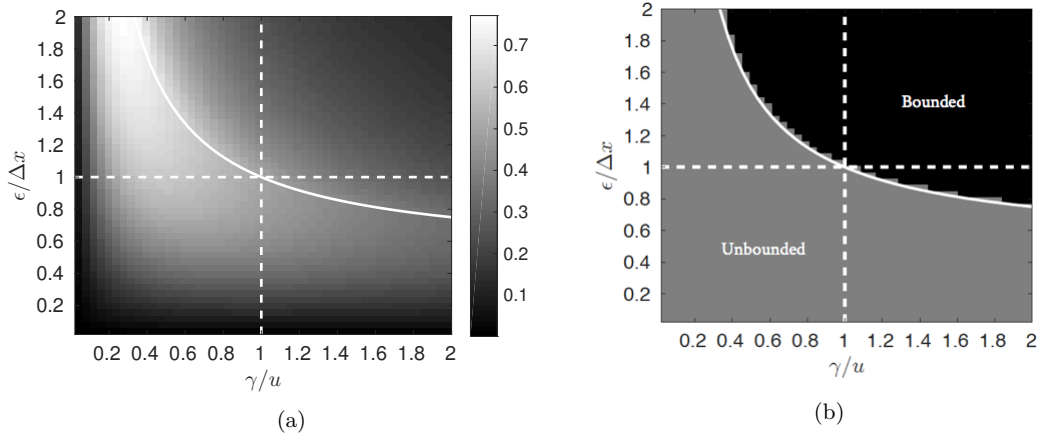

Figure 2: Diagram showing the maximum stable $CFL = u\Delta t_{mod}/\Delta x$ (a) and bounded regions (b) for 1D numerical solutions to Equation 1.

Before moving onto the next step of our analysis, we point to the tight connection between temporal stability and boundedness. While the former is a necessary condition for the latter, the stability criterion developed above implicitly used boundedness of $\phi$ by assuming bounded characteristic speeds ($\max(|1-2\phi|)$). These assumptions are appropriate as long as we seek a parameter space in which both stability and boundedness hold. Aside from its implications on numerical stability, strict boundedness of phi within $0 \leq \phi \leq 1$ is necessary because of its implication on physical realizability of the results. When coupled with the Navier-Stokes equation, values of $\phi$ out of this bound imply unphysical fluid density overshoots and undershoots. Particularly, for systems with high-density ratios, even slight deviations from the bounds can results in negative local densities.

Next, we will investigate the non-dimensional phase field parameters, $\epsilon/\Delta x$ and $\gamma/u$ to identify the subdomains that lead to bounded solutions for $\phi$ in the range $0 \leq \phi \leq 1$ considering the developed temporal stability criterion in Equation 8. Before discussing a boundedness proof, we present results of such investigation through many numerical tests. Figure 2b shows the outcome of simulations over a wide range of input parameters indicating whether a system with bounded initial condition has remained bounded. It is remarkable that the bounded versus unbounded outcomes are separated exactly by the same crossover line that allows for the maximum stable time step, as derived in Equation 9. In other words, for cases in which the stiffness in Equation 1 is dominated by the advective terms, a bounded solution in the range $0 \leq \phi \leq 1$ is not guaranteed. These solutions correspond to data below the crossover line. In contrast, when the input parameters are selected from the space above the crossover line, the numerical solution has remained bounded at all times. The results presented in Figure 2 indicate trade-offs between boundedness, numerical stiffness, and spatial resolution. While the data points above the crossover line result in strictly bounded solutions, using excessively large values of $\epsilon/\Delta x$ would lead to thick phase interfaces, and thus the effective resolution in capturing physical features such as drops and bubbles, would be suppressed. On the other hand, excessively large values of $\gamma/u$ would limit the allowable CFL leading to significant slowdown of simulations. Therefore, the crossover line (Equation 9) represents the optimal points in this three-way tradeoff. On the line itself, there is a tradeoff between spatial resolution and temporal stiffness that may be decided depending on the specific physical problem at hand.

2.2. Boundedness analysis

We adopt the same staggered finite difference spatial discretization as in [1]. This grid configuration is different from the semi-staggered grid used in [13]. Velocity fields are stored on the faces, while $\phi$ values are stored at cell centers. Equation 1 is thereby numerically approximated by

- Compute normal vectors at the cell centers
- Compute nonlinear sharpening term at cell centers then interpolate onto faces to obtain fluxes
- Compute diffusive flux on faces
- Compute advective flux after interpolating $\phi$ onto faces
- Use the divergence of the computed fluxes at time-step $k$ to update $\phi$ for time-step $k+1$,

where all derivatives are numerically computed with central differences. It is necessary to note that in [1], $\phi$ is transported after computing the advective flux with non-linear TVD upwind schemes. Afterwards, they use central differences exactly as we do in the reinitialization step to compute sharpening and diffusive fluxes and update the phase field in pseudo-time until convergence. An important caveat here is that the authors in [1] fix the normal vector field ($\vec{n}$) during reinitialization. This leads to artificial movement of the interface as the diffusive and sharpening (compressive) fluxes cannot balance each other. This problem is addressed by the same authors in [2] by projecting the diffusive fluxes in the direction defined by the fixed normal vector computed at the beginning of reinitialization. We avoid these problems by combining phase transport and reinitialization into a phase field (Equation 1), first introduced in [13]. In [13], expensive upwind schemes were used to obtain all fluxes. However, we not only use central differences for computation of diffusive and nonlinear sharpening fluxes as in [1], but we also discretize the advective flux with central differences. This is only permitted by carefully choosing $\epsilon$ and $\gamma$ to guarantee boundedness of $\phi$. Boundedness analysis on Equation 1 will be first performed for explicit Euler time integration, and then we will discuss other explicit schemes in section 2.3.

*2.2.1. Boundedness in one-dimensional flow*

Following the steps explained above, for an incompressible flow on a uniform 1D mesh, explicit Euler time stepping applied to Equation 1 results in the following update rule:

$$\phi_i^{k+1} = \phi_i^k + \Delta t (\underbrace{-u\frac{\phi_{i+1}^k - \phi_{i-1}^k}{2\Delta x}}_{\text{Advection}} + \underbrace{\gamma\epsilon\frac{\phi_{i+1}^k - 2\phi_i^k + \phi_{i-1}^k}{\Delta x^2}}_{\text{Diffusion}} + \underbrace{\gamma\frac{((\phi_{i+1}^k)^2 - \phi_{i+1}^k)n_{i+1}^{\hat{k}} - ((\phi_{i-1}^k)^2 + \phi_{i-1}^k)n_{i-1}^{\hat{k}}}{2\Delta x}}_{\text{Sharpening}}), \tag{11}$$

where $n_i^{\hat{k}} = (\phi_{i+1}^k - \phi_{i-1}^k)/\left|\phi_{i+1}^k - \phi_{i-1}^k\right| = \text{sgn}(\phi_{i+1}^k - \phi_{i-1}^k)$, and all the terms that are evaluated at time step $k$ have been moved to the right hand side. For the sake of tractability, we define two important non-dimensional parameters given by $A = \Delta t\gamma\epsilon/\Delta x^2$, which is effectively the diffusive CFL number and $B = \Delta x/(2\epsilon)$, a non-dimensional mesh size. We can thus rewrite Equation 11 in the following form:

$$\phi_i^{k+1} = \phi_i^k + AB\frac{u}{\gamma}(-\phi_{i+1}^k + \phi_{i-1}^k) + A(\phi_{i+1}^k - 2\phi_i^k + \phi_{i-1}^k) + AB(((\phi_{i+1}^k)^2 - \phi_{i+1}^k)n_{i+1}^{\hat{k}} - ((\phi_{i-1}^k)^2 + \phi_{i-1}^k)n_{i-1}^{\hat{k}}) \tag{12}$$

which after rearrangement and dropping the superscript for time-step $k$ yields the general update rule given by

$$\phi_i^{k+1} = A[1 + B(n_{i-1}(1 - \phi_{i-1}) + \frac{u}{\gamma})]\phi_{i-1} + [1 - 2A]\phi_i + A[1 + B(-n_{i+1}(1 - \phi_{i+1}) - \frac{u}{\gamma})]\phi_{i+1}$$
$$= C_{i-1}\phi_{i-1} + C_i\phi_i + C_{i+1}\phi_{i+1}. \tag{13}$$

In Figure 2b, we observed that $\phi$ remained bounded during our numerical simulations when $\epsilon/\Delta x$ and $\gamma/u$ were chosen to be above the crossover line given by Equation 9. Now, we seek to prove this observation analytically. Above the line, $\epsilon/\Delta x \geq (u/\gamma + 1)/2$ and we know from Equation 8 that we require $\Delta t \leq \Delta x^2/(2\gamma\epsilon)$ for numerical stability. This results in $A \leq 1/2$ and $B(1 + u/\gamma) \leq 1$. Subsequently, in Equation 13 we can show that if for a given time-step $\phi$ is bounded($0 \leq \phi_i \leq 1$ for all $i$) the coefficients $C_{i-1}$, $C_i$ and $C_{i+1}$ are all positive. While $C_i = 1 - 2A \geq 0$ requires no explanation, for $C_{i-1}$ and $C_{i+1}$ we respectively have

$$\frac{C_{i-1}}{A} = 1 + B(n_{i-1}(1-\phi_{i-1}) + \frac{u}{\gamma}) \geq 1 + B(-1 + \frac{u}{\gamma}) \geq 1 - B(1 + \frac{u}{\gamma}) \geq 0, \tag{14}$$

and

$$\frac{C_{i+1}}{A} = 1 + B(-n_{i+1}(1-\phi_{i+1}) - \frac{u}{\gamma}) \geq 1 - B(1 + \frac{u}{\gamma}) \geq 0. \tag{15}$$

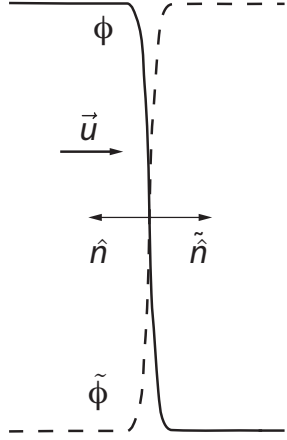

Figure 3: An illustration of the transformation used to introduce $\tilde{\phi}$ for analyzing the upper bound to $\phi$.

This means that for simulation parameters above the crossover line and $\Delta t \leq \Delta t_{mod}$, if the phase field is bounded at a given time-step ($0 \leq \phi_i^k \leq 1$ for all $i$), then for the next time-step we have $\phi_i^{k+1} \geq 0$ for all $i$ values.

To prove boundedness, the remaining task is to demonstrate that $\phi_i^{k+1} \leq 1$ for all $i$ values. We define auxiliary variables $\tilde{\phi} \equiv 1 - \phi$, and $\tilde{n} = \nabla\tilde{\phi}/\left|\nabla\tilde{\phi}\right| = -\hat{n}$ as in Figure 3. Discretely, if we substitute $1 - \tilde{\phi}_i^k = \phi_i^k$ into Equation 11, the update rule for $\tilde{\phi}$ is found to be exactly the same as $\phi$:

$$\tilde{\phi}_i^{k+1} = \tilde{\phi}_i^k + \Delta t \left( -u\frac{\tilde{\phi}_{i+1}^k - \tilde{\phi}_{i-1}^k}{2\Delta x} + \gamma\epsilon\frac{\tilde{\phi}_{i+1}^k - 2\tilde{\phi}_i^k + \tilde{\phi}_{i-1}^k}{\Delta x^2} + \gamma\frac{[(\tilde{\phi}_{i+1}^k)^2 - \tilde{\phi}_{i+1}^k]\tilde{n}_{i+1}^k - [(\tilde{\phi}_{i-1}^k)^2 + \tilde{\phi}_{i-1}^k]\tilde{n}_{i-1}^k}{2\Delta x} \right). \tag{16}$$

Therefore, the proof for positivity of $\phi$ applies for $\tilde{\phi}$ as well, which translates to $\phi \leq 1$. With that, we have finished proving our observation from Figure 2b that in a 1D flow with stable Euler time-integration if $\phi$ is initially bounded between 0 and 1 our scheme will preserve those bounds as long as the free parameters $\epsilon/\Delta x$ and $\gamma/u$ are chosen above the crossover line defined in Equation 9.

Before moving on to multi-dimensional flows and higher order time-schemes, we would like to explain why boundedness is not guaranteed under the crossover line when $\Delta t$ is chosen according to Equation 8. Considering the definition of $A$ and $B$, strictly under the crossover line we have $A < 1/2$ and $B(1+u/\gamma) > 1$. As a result, while $C_i$ in Equation 13 is positive, $C_{i+1}$ can be negative, especially where $\hat{n}$ and $u$ are of the same sign. This is problematic when $r = \phi_{i+1}^k/\phi_i^k \gg 1$ which happens often in regions of transition from bulk phase to interfacial zones, as shown in Figure 4. Despite this, for the droplet advection problem, it is possible to extend the boundedness region slightly below the crossover line by adaptively reducing $\Delta t$ and thus $|C_{i-1}|$ and $|C_{i+1}|$ as suggested by Equation 13. This is shown in Figure 5a. For simulations using parameters from above the crossover line, $\Delta t_{mod}$ from Equation 8 guarantees boundedness. Conversely, for parameters from below the crossover line, adaptive time-steps much smaller than $\Delta t_{mod}$ are required to maintain boundedness of $\phi$, as shown in Figure 5b. The reported CFL numbers in this Figure reveal the significant penalty on temporal resolution required to push the bounded region below the crossover line. We therefore conclude that in practice only $\epsilon/\Delta x$ and $\gamma/u$ pairs on or above the crossover line should be used for the advection of the interface using this diffuse interface method. Figure 5c exemplifies the discrete conservation in our proposed diffuse interface method and in Figure 5d, the region just above the crossover line had the smallest total shape error (defined and discussed in length in Section 3.1).

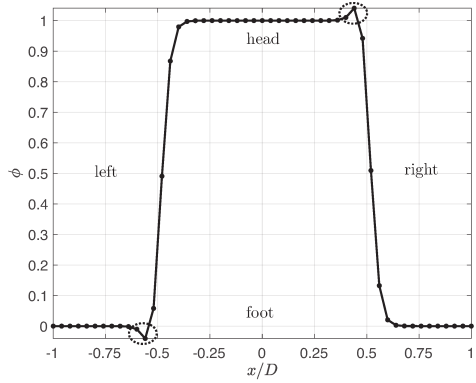

Figure 4: When $(\epsilon/\Delta x, \gamma/u)$ is below the crossover line, undershoots and overshoots can be observed in regions where $\phi$ is close to 0 or 1 and the advection and sharpening terms of Equation 11 have matching signs (e.g., see the left foot and right head).

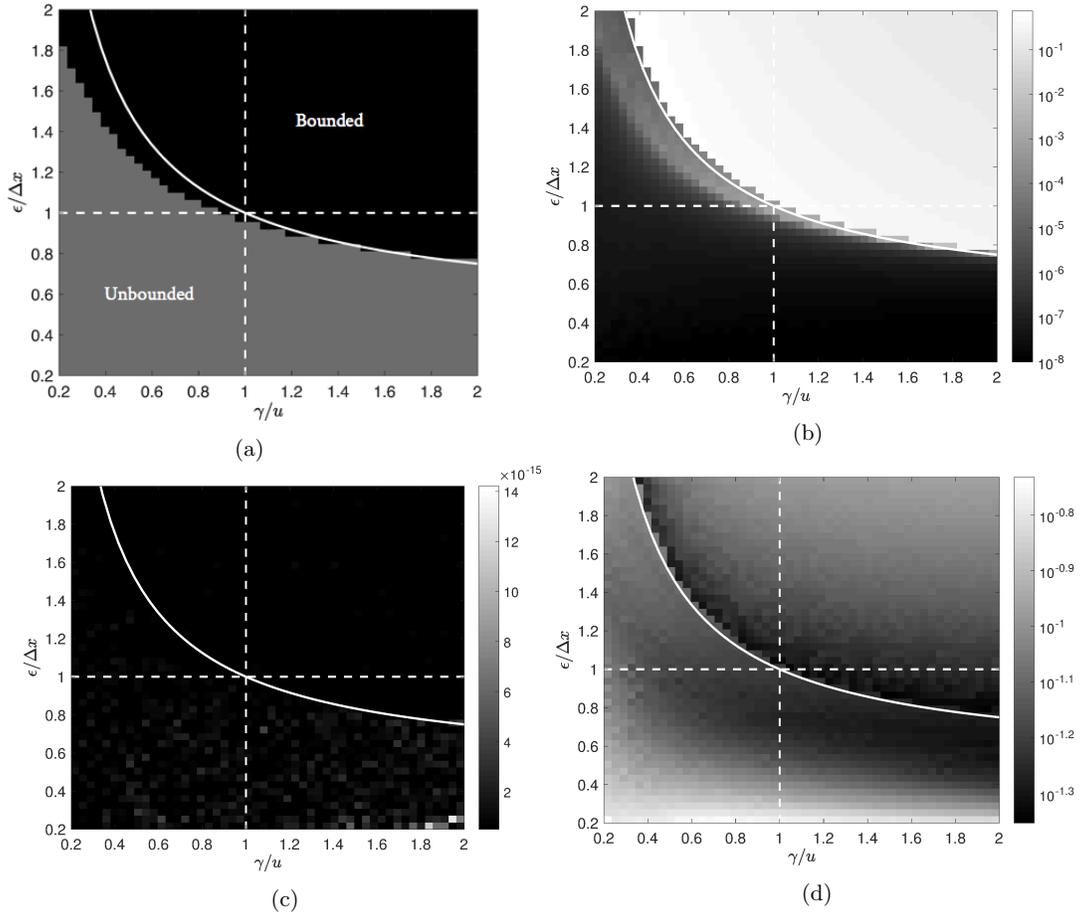

Figure 5: Boundedness (a), $CFL$ (b), conservation (c) and $Err_{tot}$ (d) using $\Delta t = \Delta t_{mod}$ above crossover line and an adaptively chosen time-step below the crossover line. Results are from the 1D drop advection problem for one period for many $\epsilon, \gamma$ values with $nx = 51$.

Now we will extend our analysis to variable velocity incompressible flows in higher dimensions.

*2.2.2. Boundedness for General Incompressible Flow Fields in Higher Dimensions*

For a general flow in 2D or 3D on a uniform isotropic mesh, the non-dimensional parameters $\epsilon/\Delta x$ and $\gamma/|\vec{u}|_{max}$ will determine the boundedness of the solution to Equation 1. Similar to Equation 9, we will demonstrate that if we select these parameters such that:

$$\epsilon/\Delta x \geq \frac{\gamma/|\vec{u}|_{max} + 1}{2\gamma/|\vec{u}|_{max}}, \qquad (17)$$

then boundedness of $\phi$ is guaranteed. Given that we pick the simulation parameters from above the crossover line, now defined by Equation 17, the time-step for stability is limited by diffusion such that $\Delta t \leq \Delta x^2/(2D_f\gamma\epsilon)$, where $D_f$ is the number of spatial dimensions. Here we will skip linear stability analysis of the multi-dimensional problem as presented in section 2.1. Instead, we will show that the diffusion limited time-step suggested above guarantees boundedness, which is a sufficient condition for stability.

With the new diffusive CFL, $A = \Delta t\gamma\epsilon/\Delta x^2$, we see that $A \leq 1/(2D_f)$. Additionally, maintaining definition of $B = \Delta x/(2\epsilon)$, one can see from Equation 17 that $B(1 + |\vec{u}|_{max}/\gamma) \leq 1$. Next, with these inequalities at hand, we proceed with proof of boundedness for a 2D scenario, from which extension to 3D will be straight-forward. After rearranging the discretization of Equation 1 for an arbitrary non-uniform incompressible flow in a manner similar to Equation 13, we have:

$$\phi_{i,j}^{k+1} = C_{i-1,j}\phi_{i-1,j} + C_{i,j-1}\phi_{i,j-1} + C_{i,j+1}\phi_{i,j+1} + C_{i+1,j}\phi_{i+1,j} + C_{i,j}\phi_{i,j}. \tag{18}$$

The coefficient multiplying $\phi_{i,j}$ in Equation 18 is given by

$$C_{i,j} = 1 - 4A - \frac{BA}{\gamma}(u_{i+\frac{1}{2},j} - u_{i-\frac{1}{2},j} + v_{i,j+\frac{1}{2}} - v_{i,j-\frac{1}{2}}), \tag{19}$$

where the last term vanishes due to incompressibility, and given that $A \leq \frac{1}{7}(2D_f) = 1/4$, one can conclude

$$C_{i,j} \geq 0. \tag{20}$$

The rest of the coefficients multiplying neighboring cell $\phi$ values are also positive as

$$\frac{C_{i-1,j}}{A} = 1 + B(n_{x,i-1,j}(1 - \phi_{i-1,j}) + \frac{u_{i-1/2,j}}{\gamma}) \geq 1 + B(-1 + \frac{u_{i-1/2,j}}{\gamma}) \geq 1 - B(1 + \frac{|\vec{u}|_{max}}{\gamma}) \geq 0, \tag{21}$$

and

$$\frac{C_{i+1,j}}{A} = 1 + B(-n_{x,i+1,j}(1 - \phi_{i+1,j}) - \frac{u_{i+1/2,j}}{\gamma}) \geq 1 + B(-1 - \frac{u_{i+1/2,j}}{\gamma}) \geq 1 - B(1 + \frac{|\vec{u}|_{max}}{\gamma}) \geq 0. \tag{22}$$

Similarly, it is easy to verify that $C_{i,j-1}$ and $C_{i,j+1}$ are also positive. Given that all coefficients in Equation 18 are positive, starting with a bounded initial $\phi$, we conclude that the phase field will remain positive in the next time-step. To prove boundedness, we just need to demonstrate that $\phi$ will also remain below 1. This is straight-forward by using the same technique as in 1D by defining $\tilde{\phi} \equiv 1 - \phi$, and verifying that it satisfies the same discrete evolution equation as $\phi$ for an incompressible flow.

With careful inspection of Equations 21 and 22, it is apparent that the inequalities are less tight than those presented for 1D flow in 14 and 15. This observation is three-fold: first, by definition $x$ and $y$ components of the normal vector have a magnitude smaller than 1, second, $u$ and $v$ are smaller than $|\vec{u}|$, and third $|\vec{u}| \leq |\vec{u}|_{max}$ everywhere in the domain. As a matter of fact, in 2D, one can achieve boundedness via a less restrictive criterion given by

$$\epsilon/\Delta x \geq \frac{\gamma/\max(u_{max}, v_{max}) + 1}{2\gamma/\max(u_{max}, v_{max})}, \tag{23}$$

and similarly for 3D,

$$\epsilon/\Delta x \geq \frac{\gamma/\max(u_{max}, v_{max}, w_{max}) + 1}{2\gamma/\max(u_{max}, v_{max}, w_{max})}, \tag{24}$$

guarantees boundedness. In the remainder of this paper, however, we will use the general condition presented in Equation 17 for boundedness, since in most realistic flows the conditions given by Equations 23 or 24 are practically too close to Equation 17 to offer a significant advantage. Moreover, it should be noted that the crossover line condition must be satisfied at all times. Therefore, in settings in which $|\vec{u}|_{max}$ grows in time, $\gamma$ should be adapted appropriately.

*2.3. Higher order explicit time-stepping schemes*

It is straight-forward to predict how the recommendations developed thus far on time-step size, and parameters $\epsilon/\Delta x$ and $\gamma/|\vec{u}|_{max}$ can be extended to discretizations with higher order temporal accuracy. Even though we were not able to extend our formal proof from section 2.2.2, our numerical examinations over a wide range of multi-dimensional settings confirm that Equation 17 and maximum stable diffusive CFL are sufficient criteria for boundedness of $\phi$ when higher order Runge-Kutta (RK) methods are used. While we will present example calculations in section 3.2, here we briefly provide intuitive explanations regarding stability and boundedness when higher order time-advancement schemes are used.

Based on our detailed examinations of the explicit Euler time-advancement scheme, we have made two observations so far:

Firstly, as long as the parameters $\epsilon/\Delta x$ and $\gamma/|\vec{u}|_{max}$ are selected from above the cross-over line (Equation 17), the stiffness of the system is dominated by the diffusion term, and temporal stability is controlled by a diffusive CFL. This conclusion can be extended to higher order RK schemes. This is due to the fact that among RK methods, explicit Euler is the least accommodative scheme for advective operators, yet the time-step is diffusion-limited when Equation 17 is satisfied. Higher order RK schemes significantly relieve restrictions on time-step due to advective operators without as much relief on restrictions by diffusive operators. Consistent with this explanation, we find stable solutions when the maximum time-step is given by:

$$\Delta t_{mod}^{Eul.} = \frac{\Delta x^2}{2D_f\gamma\epsilon}, \tag{25}$$

$$\Delta t_{mod}^{RK2} = \frac{\Delta x^2}{2D_f\gamma\epsilon}, \tag{26}$$

$$\Delta t_{mod}^{RK3} = \frac{2.51\Delta x^2}{4D_f\gamma\epsilon}, \text{ and} \tag{27}$$

$$\Delta t_{mod}^{RK4} = \frac{2.79\Delta x^2}{4D_f\gamma\epsilon}. \tag{28}$$

These equations respectively correspond to maximum diffusive CFL for explicit Euler, second-order explicit midpoint RK scheme (RK2), the low-storage third-order explicit RK method of Wray (RK3) [15], which is a popular time-stepping scheme for solving the incompressible Navier-Stokes equations (e.g., see [16, 14]), and the classic fourth-order explicit RK method (RK4).

The second observation from the preceding sections is that, as long as the parameters $\epsilon/\Delta x$ and $\gamma/|\vec{u}|_{max}$ are selected from above the cross-over line defined by Equation 17, the numerically evolved $\phi$ field remains bounded between zero and one. Our numerical studies confirm that this condition also holds for higher order RK schemes. To explain this intuitively, we introduce the semi-discretized format of Equation 1 as:

$$\frac{d\phi_i}{dt} = \tilde{C}_{i-1}\phi_{i-1} + \tilde{C}_i\phi_i + \tilde{C}_{i+1}\phi_{i+1}, \tag{29}$$

for a one-dimensional setting. Above the crossover line, as shown in Equations 14 and 15, we have $\tilde{C}_{i-1} = C_{i-1}\Delta t \geq 0$ and $\tilde{C}_{i+1} = C_{i+1}\Delta t \geq 0$ while $\tilde{C}_i$ is negative. Equation 29 can be viewed as a dynamical system for $\phi_i$ which analytically directs it towards its fixed point, $\phi_i \to \frac{\tilde{C}_{i-1}\phi_{i-1}+\tilde{C}_{i+1}\phi_{i+1}}{-\tilde{C}_i}$. Note that because of the signs of $\tilde{C}_{i-1}$, $\tilde{C}_i$ and $\tilde{C}_{i+1}$, $\phi_i$ should converge to a positive value based on Equation 29. In settings where the neighboring values of $\phi$ and $\tilde{C}$ coefficients change in time, one can at least conclude that $\phi_i$ cannot become negative as long as $\phi_{i+1}$ and $\phi_{i-1}$ remain bounded. Similar to our discussions in section 2.2.1, by examining the evolution equation for $\tilde{\phi} \equiv 1 - \phi$, one can show that $\phi_i$ also remains smaller than one.

These observations mean that starting with a bounded initial condition, the semi-discrete dynamical system preserves the boundedness of $\phi$. Same arguments can be made for multi-dimensional settings, similar to what has been shown in section 2.2.2. Of course, compared to explicit Euler, higher order RK schemes are able to follow time-continuous evolution of $\phi_i$ more accurately. Hence, it is not surprising that they are better in maintaining the boundedness property of the solution. We here point to the fact that Equation 17 is actually a condition for boundedness of the semi-discrete equation. Therefore, the simulation parameters should be chosen according to this criterion regardless of the RK time-stepping scheme.

Figure 6 shows the boundedness and stability phase diagrams for the drop advection problem introduced in Figure 1 using different time-stepping schemes with their corresponding $\Delta t_{mod}$ values. Our boundedness analysis for above the crossover line still holds in spite of using larger time steps. Conversely, for points under the crossover line, both unbounded and unstable behavior is observed. The points that are marked as unbounded in Figure 6 may imply that $\phi$ values outside of $[0, 1]$ were detected either at the end of a time-step or at intermediate sub-steps of an RK scheme.

It should be noted that for simulations using points from under the crossover line, we used an appropriate mixed advective-diffusive criterion similar to Equation 8 for selection of the time-step. However, as explained in Section 2.1, once the phase field goes unbounded and the assumption of $\max(|1 - 2\phi|) \leq 1$ is invalidated, our analysis for stability would not be completely accurate. That is why some unstable solutions are observed under the crossover line for the RK3 and RK4 schemes in Figure 6, despite the fact that nominal stability criteria are enforced.

Before venturing into numerical studies, we would like to revisit the developed boundedness criterion (Equation 17) in the context of the work of [1] which was the first to use an approach similar to Equation 1. Namely in their work, Equation 1 was used as a reinitialization step and after physical advection was performed at each time-step. In other words, the phase field is updated in pseudo-time until convergence using the right-hand-side terms of Equation 1 and $\vec{u} = 0$ everywhere. Therefore, their phase field reinitialization step corresponded to $\gamma/u \to \infty$. According to Equation 17, the recommended value for $\epsilon$ is $\epsilon \geq \Delta x/2$. They adopted $\epsilon = \Delta x/2$ and RK2 time-stepping, which puts them in the bounded diffuse interface regime as presented in section 2.3.

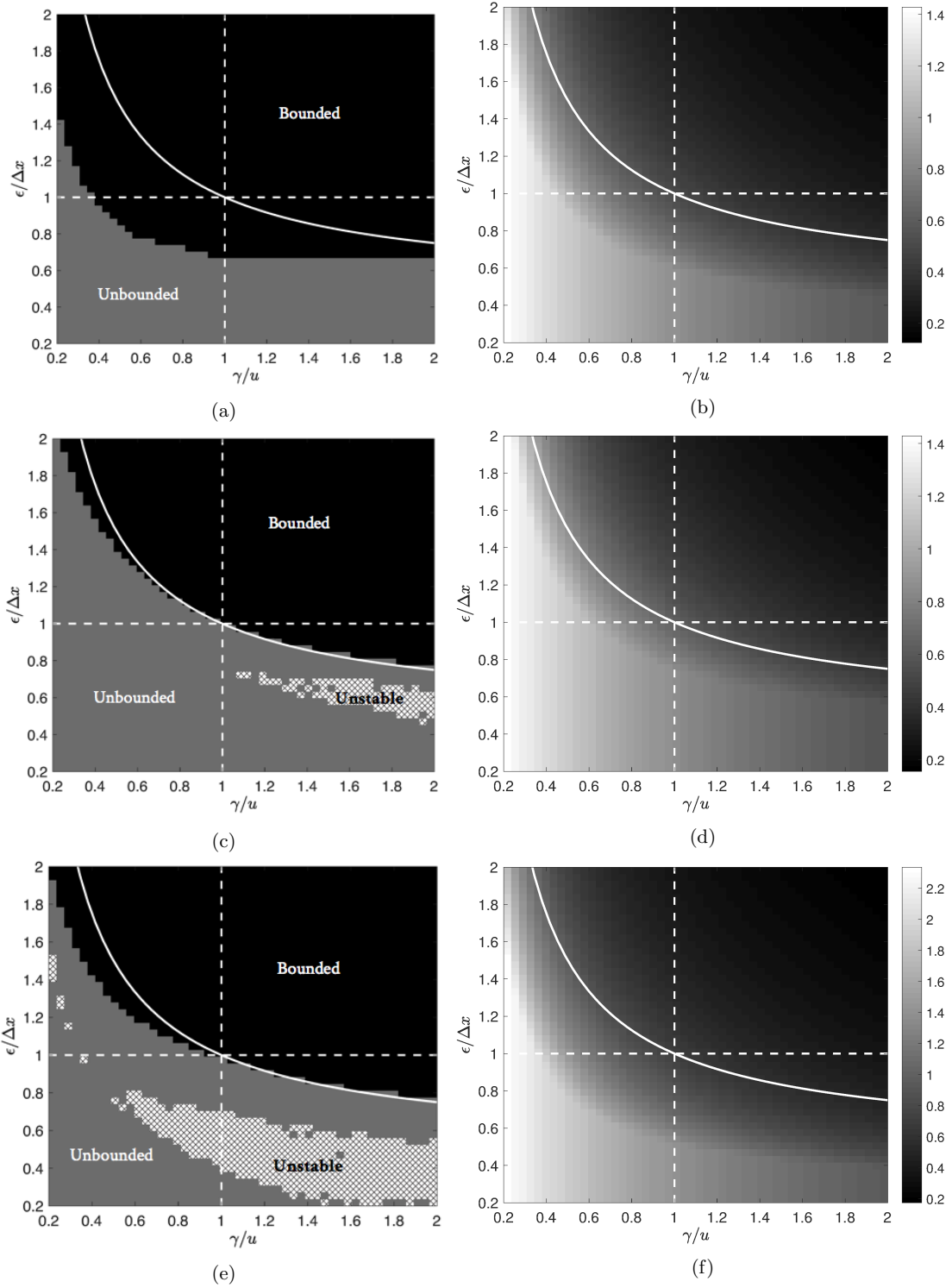

Figure 6: Boundedness (a,c,e) and $CFL$ (b,d,f) for RK2 using $\Delta t = \Delta t_{mod}^{RK2}$ (a,b), RK3 using $\Delta t = \Delta t_{mod}^{RK3}$ (c,d) and RK4 using $\Delta t = \Delta t_{mod}^{RK4}$ (e,f) after one period for all $\epsilon, \gamma$ values

## 3. Numerical studies

Based on our analysis throughout section 2, we have established that as long as the input parameters are chosen according to Equation 17, on a staggered uniform and isotropic mesh, applying the described central difference scheme for spatial derivatives results in bounded transport of $\phi$, regardless of the number of dimensions to the flow. In this section, we will verify these results via numerical tests considering multi-dimensional scenarios and non-uniform velocity fields (Section 3.2). Before that, in Section 3.1 we will use 1D numerical simulations to quantitatively assess the convergence properties of discrete solutions to Equation 1. Additionally, in Section 3.3 we provide standard test cases such as the drop in shear flow and Zalesak disk, along with a quantitative comparison against the scheme of [13]. This is to ensure that our proposed discretization retains sufficient levels of accuracy for these test cases while also guaranteeing boundedness.

We use RK4 time integration in all numerical experiments. When the simulation parameters are chosen from above the crossover line, the time-step is given by $\Delta t_{mod}^{RK4}$ (see Equation 28). As derived in section 2.1, for simulations with $\epsilon$ and $\gamma$ values below the crossover line, appropriate advective-diffusive time-steps are chosen to enforce stability.

### 3.1. Mesh refinement studies

In this section, we use our 1D drop advection test case to examine the convergence behavior of Equation 1 with respect to mesh refinement. Errors incurred by phase field methods are two fold. First, there is the error in approximating a sharp interface with a diffuse interface. We will denote this error by $E_\epsilon$. Second, there are truncation errors involved in numerically solving the phase field equation, which we will denote by $E_\Delta$. The reason for this naming is that the discretization error can be eliminated by refining the space-time mesh and the error associated with the diffuse approximation can be eliminated by refining $\epsilon$, subject to the boundedness condition. We will study the convergence behavior of both of these errors and explain how $\epsilon$ and $\gamma$ should be altered to obtain optimal convergence behavior. The specific problem considered here is advection of a 1D drop by precisely one period in a periodic domain. The exact solution to the governing PDE is $\phi(x,T) = \phi(x,0)$, where the periodic domain is $-1 \leq x \leq 1$, as shown in Figure 1. Advection speed is $u = 1$ and the time period is $T = 2$. In the sharp interface limit, this solution is a top-hat function which we denote as $\phi = TH(x)$. The initial condition to each problem is a top-hat profile relaxed to the steady solution of the interfacial profile, without the advective term present. Therefore, we can measure $E_\epsilon$ as the difference between the top-hat profile and the initial condition, $E_\epsilon = \|\phi(x,0) - TH(x)\|_1$, and $E_\Delta$ as the difference between the initial condition and the one-period advected profile $E_\Delta = \|\phi(x,T) - \phi(x,0)\|_1$. The total error is defined as

$$E_{tot} = \|\phi(x,T) - TH(x)\|_1 \leq E_\epsilon + E_\Delta. \tag{30}$$

By fixing the values of $\epsilon$ and $\gamma$, and refining $\Delta x$, one can study the convergence of $E_\Delta$ and $E_\epsilon$ with respect to mesh refinement (see Figure 7a). Similarly, we can vary $\epsilon$ while keeping the mesh and $\gamma$ constant, as shown in Figure 7b. Varying $\gamma$ while maintaining the values of $\epsilon$ and $\Delta x$ results in no significant change in the error as shown in Figure 7c. However, it is apparent that $E_\Delta$ and $E_\epsilon$ have first order dependence on $\Delta x$ and $\epsilon$ respectively, while their cross-dependence is relatively negligible. Ideally, using second-order central differences should yield second-order convergence of the truncation error, $E_\Delta \sim \Delta x^2$; however, in this 1D test case, in regions where the normal vector ($\hat{n}$) switches sign, the PDE involves a weak singularity that can be eliminated in the limit of analytical solutions for a thin interface (when switch happens at $\phi = 0$ or 1). Accumulation of errors due to this effect brings down the convergence rate to $E_\Delta \sim \Delta x$. By conducting numerical tests at different regions of the $\epsilon - \gamma - \Delta x$ parameter space, we confirmed that these conclusions are maintained as long as the stated stability and boundedness conditions are met.

To achieve optimal convergence behavior, $E_\Delta$ and $E_\epsilon$ should be of the same order. Hence, it is reasonable to choose $\epsilon \sim \Delta x$. This means that for a grid convergence study, in addition to using smaller mesh spacing, $\epsilon$ should also be reduced proportional with $\Delta x$. Therefore, maintaining a fixed point in $\epsilon/\Delta x - \gamma/u$ parameter space above the crossover line would be an appropriate strategy for optimal convergence, as confirmed by the results presented in Figure 8. Since for a fixed mesh, the total error scales with $\epsilon$, we should chose a point on the crossover line to minimize the error. This is illustrated in Figure 9, as at all resolutions, the errors from simulating with input parameters on the crossover line(shown with '×') are clearly smaller than errors when the same $\gamma/u$ is used along with a larger $\epsilon/\Delta x$ from above the crossover line(shown with '+'). Figure 9 shows the expected first order convergence rate of the total error for these two and three other $\epsilon/\Delta x - \gamma/u$ pairs chosen from the crossover line as a function of mesh spacing. In general, from Figure 9 we can deduce that setting $\epsilon/\Delta x$ and $\gamma/|\vec{u}|_{max}$ to both be $O(1)$ and on the crossover line provides small errors with reasonable trade-off between stiffness, $E_\Delta$ and $E_\epsilon$, while we expect the specific optimal numbers may be problem dependent.

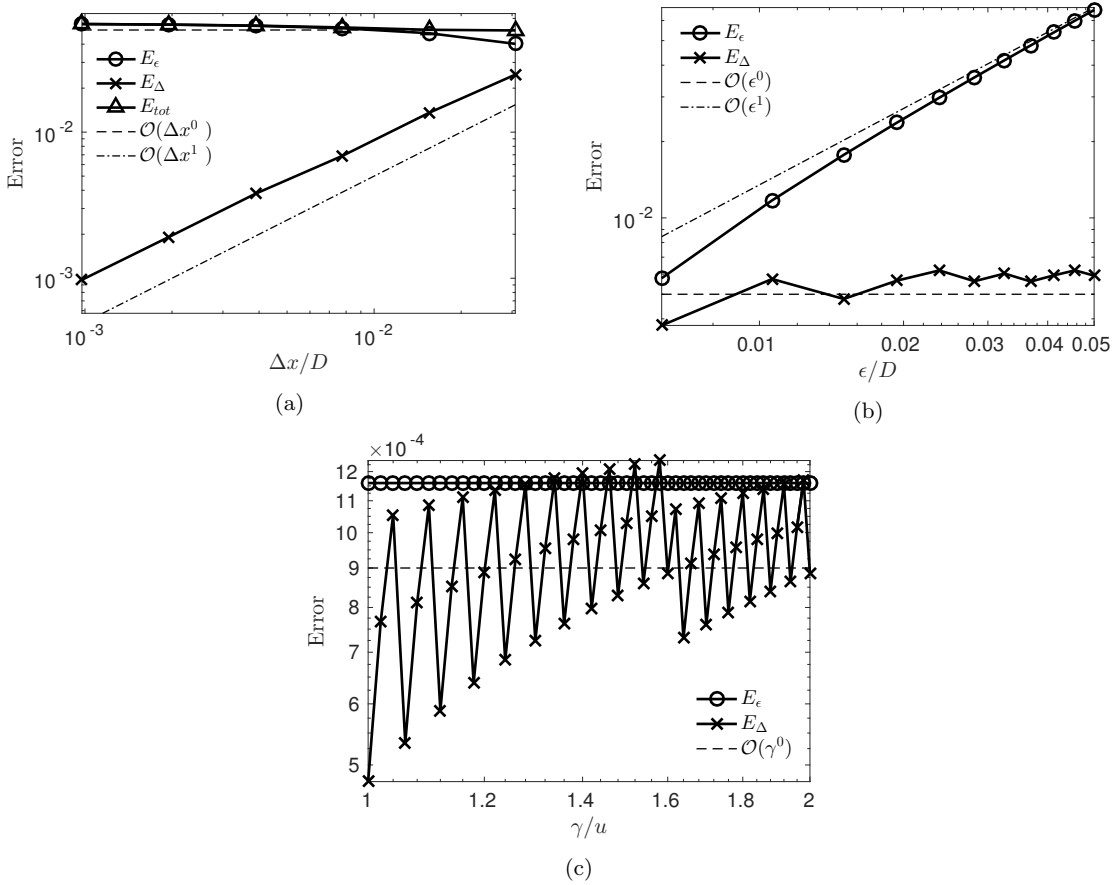

Figure 7: Comparison of different errors for the 1D drop advection problem with drop size $D = 1$ using RK4 time-stepping. Presented are cases with constant $\epsilon = 1/25$ and $\gamma/u = 1$, while varying $\Delta x$(a), constant $\Delta x = 1/320$ and $\gamma/u = 1$, while varying $\epsilon$(b), and constant $\epsilon = 1/1600$ and $\Delta x = 1/1600$, while varying $\gamma$

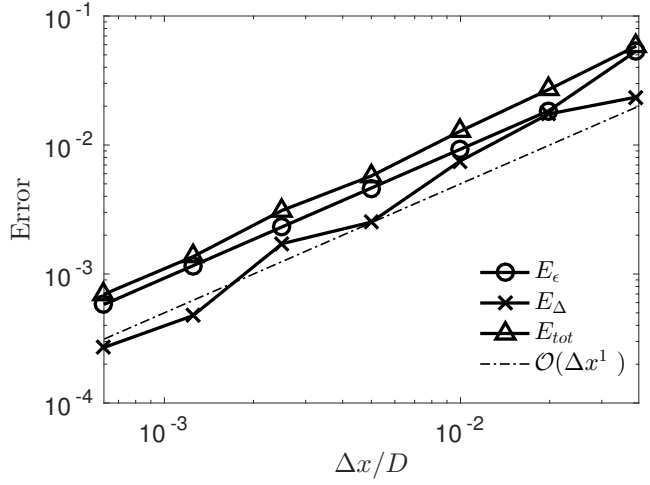

Figure 8: $E_\epsilon$, $E_\Delta$ and $E_{tot}$ as a function of $\Delta x/D$ for 1D drop advection test with $\epsilon/\Delta = \gamma/u = 1$.

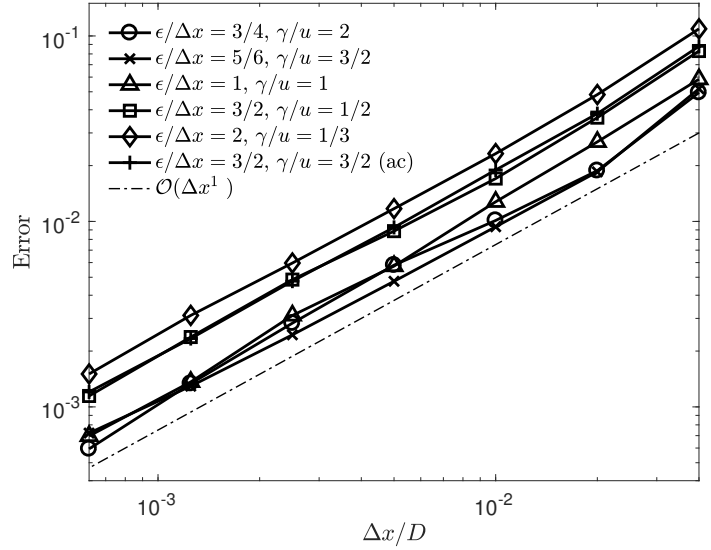

Figure 9: $Err_{tot}$ from the 1D drop advection test as a function of $\Delta x/D$ for selected $\epsilon/\Delta x$-$\gamma/u$ pairs on and above the crossover line (ac).

*3.2. Numerical boundedness in two-dimensional domains*

Advection of a 2D drop for two different test cases on isotropic, uniform meshes is studied in this section. In the first test (Section 3.2.1), a drop is advected with uniform velocity, and in the second test (Section 3.2.2), the drop is advected by a spatially random divergence-free velocity field. We study the boundedness of the 2D solution in the $\epsilon/\Delta x - \gamma/u$ parameter space by performing simulations on seven different pairs of $\epsilon/\Delta x$ and $\gamma/u$ values for each test case as indicated in table 2. Above the crossover line defined in Equation 17, when using RK4 time-advancement we consider

$$\Delta t_{mod}^{RK4,2D} = \frac{2.79\Delta x^2}{8\gamma\epsilon}. \tag{31}$$

Since some of the input points are below the crossover line, we used a mixed advective-diffusive criterion for selection of the time step at those points given by:

$$\Delta t_{mod}^{RK4,2D,bc} = \frac{2.83\Delta x}{\max(u) + \max(v) + 2\gamma}. \tag{32}$$

*3.2.1. 2D uniform velocity field*

A drop with diameter $D = 1$ is initialized at the center of a $2 \times 2$ periodic box and advected with velocity $(u, v) = (\sqrt{2}/2, \sqrt{2}/2)$. At time $t = T = 2\sqrt{2}$, the drop returns to its initial position. In Table 2, contour levels of $0-\varepsilon$, $0.5$ and $1+\varepsilon$, where $\varepsilon$ is machine precision, are plotted at $t = T$ along with the initial condition for seven different $\epsilon/\Delta x$ and $\gamma/u$ values. The $\phi$ profiles over the $x$ plane at $y = 1$ are also included. The simulations are performed with $\Delta x = D/25$. The solution for a pair of $\epsilon/\Delta x$ and $\gamma/u$ values is bounded at time T if in addition to the initial condition, only one contour line corresponding to $\phi(T) = 0.5$ is visible. Moreover, if at any time during a simulation any point in space acquires a value for $\phi$ outside of the $[0, 1]$ interval, the case will be declared unbounded. From Table 2, it is clear that only points 6 and 7 result in $\phi$ values below 0 and above 1 (shown with × in the $\epsilon/\Delta x - \gamma/u$ phase diagram), whereas points 1-5 remain bounded (shown with ◯ in $\epsilon/\Delta x - \gamma/u$ phase diagram). It is of no surprise to see that the bounded region is extended below the crossover line defined by Equation 17 and featured in Table 2. As discussed in Section 2.2.2, the less conservative crossover line defined in 23 lies below point 5 but above points 6 and 7, explaining why points 6 and 7 are the only unbounded cases.

Accuracy can be qualitatively assessed by comparing the $\phi = 0.5$ contour to the initial condition and the closeness of the $\phi$ profiles to sharp interface drop. However, it is difficult to compare the accuracies amongst points, so we include the norm-1 errors for the seven points in Table 1. The observed general trends are consistent with the discussion presented in Section 3.1.

| Point | $E_\Delta$ | $E_\epsilon$ | $E_{tot}$ |
|---|---|---|---|
| 1 | 0.007625 | 0.06380 | 0.06450 |
| 2 | 0.001775 | 0.04335 | 0.04350 |
| 3 | 0.01040 | 0.03638 | 0.03675 |
| 4 | 0.002600 | 0.06380 | 0.06430 |
| 5 | 0.005725 | 0.03918 | 0.03953 |
| 6 | 0.002750 | 0.03510 | 0.03535 |
| 7 | 0.008425 | 0.03088 | 0.03198 |

Table 1: $E_\epsilon$, $E_\Delta$ and $E_{tot}$ for the 2D drop advection problem with a uniform velocity using seven $\epsilon/\Delta x - \gamma/u$ pairs.

*3.2.2. 2D random velocity field*

In this final study, we subject our analysis to a strenuous test in which time-invariant velocity values could change significantly from cell to cell while we merely control the maximum velocity magnitude and enforce incompressibility. We assess boundedness at all times and at all points in the domain, and the slightest undershoot or overshoot observed would disqualify the simulation. Specifically, in this test a drop with diameter $D = 1$ is initialized at the center of a $2 \times 2$ box with slip boundary conditions. The velocity field, which does not vary with time, is computed by first independently assigning each mesh point a random velocity magnitude and direction from independent uniform distributions. Then we make the velocity field solenoidal by solving a Poisson equation and projecting the solution. Finally, we scale the velocity values such that $|\vec{u}|_{max} = 1$. The phase field is integrated for $T = 10$ and inspected at all times and mesh points for boundedness. We perform this test in the $\epsilon/\Delta x - \gamma/|\vec{u}|_{max}$ phase-space using a relatively coarse mesh with $\Delta x = D/25$. Table 2 shows how severely the drop and $\phi$ profiles are affected using this random velocity field. Nevertheless, the phase field remains bounded for all $\epsilon/\Delta x - \gamma/u$ pairs except point 7 (Table 2). With that we confirm that the criteria given in 17 is a sufficient and conservative requirement for boundedness. Moreover, the random velocity field is seemingly more forgiving in terms of boundedness, compared to the uniform velocity test case. This can be mainly attributed to the fact that for the random field, the maximum velocity locations in the domain are less likely to fall on an interface and in the direction normal to it.

| | uniform velocity | | random velocity | |
|---|---|---|---|---|
| Boundedness | | | | |
| Case | $\phi$ contour | $x$ slice | $\phi$ contour | $x$ slice |
| 1: $\epsilon/\Delta x = 3/2$, $\gamma/u = 1/2$ | | | | |
| 2: $\epsilon/\Delta x = \gamma/u = 1$ | | | | |
| 3: $\epsilon/\Delta x = 5/6$, $\gamma/u = 3/2$ | | | | |
| 4: $\epsilon/\Delta x = \gamma/u = 3/2$ | | | | |
| 5: $\epsilon/\Delta x = \gamma/u = 9/10$ | | | | |
| 6: $\epsilon/\Delta x = \gamma/u = 4/5$ | | | | |
| 7: $\epsilon/\Delta x = \gamma/u = 7/10$ | | | | |

Table 2: 2D drop advection in a uniform (left) and random (right) velocity field: For each numerical experiment, bounded ('o') and unbounded ('×') cases are marked in the first row, while the rows below show contours of $\phi = 0 - \varepsilon$, 0.5 and $1+\varepsilon$ as well as a cross-section of the domain showing $\phi$ profiles at $t = T$.

*3.3. Standard accuracy tests*
*3.3.1. Drop in shear flow*

In this standard test for interface capturing methods, a 2D drop is placed in a vortex which periodically varies in time. Specifically, a drop of radius $R = 0.15$ is centered at (0.5; 0.75) inside a unit sized box, and the velocity field is given by the stream function,

$$\Psi(x,t) = \frac{1}{\pi} sin^2(\pi x) sin^2(\pi y) cos(\frac{\pi t}{T}), \tag{33}$$

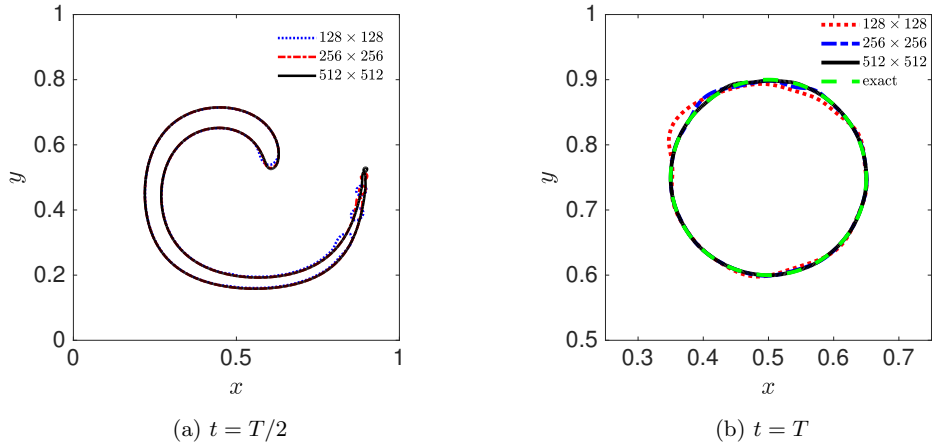

(a) $t = T/2$          (b) $t = T$

Figure 10: Interfacial profiles at $t = T/2$ for drop in deformation field. Increasing resolution from left to right and top to bottom results in an intact ligament which does not break into drops at $t = T/2$ (a). Interfacial profiles at $t = T$ for drop in deformation field plotted against initial drop. The final solution and initial drop should lay on top of each other with a perfect two-phase flow numerical method (b).

where the period of the imposed velocity field is $T = 4$. Due to the periodic nature of the velocity, this test case has an exact solution at $t = T$, which is given by the initial conditions, or in other words, $\phi_{exact}(t = T) = \phi(t = 0)$. RK4 is used for time-integration with a fixed time step of $dt = \Delta x/8$. The ability of the scheme at capturing the thin elongated drop at $t = T/2$ gives a qualitative measure of accuracy. Moreover, by comparing the final solution with the initial conditions, the accuracy of the interface capturing scheme is quantitatively measured. For convergence study, we increase the mesh resolution while keeping $\epsilon/\Delta x = 0.7$ and $\gamma/|\vec{u}|_{max} = 2.5$ fixed.

In Figure 10a, the state of the drop at $t = T/2$ can be seen for different resolution simulations. The drop is stretched by the vortex field and artificial drops appear at low resolutions, which are eliminated as we increase the resolution. In Figure 10b, the interfacial profiles at the end of the simulation (t=T) are plotted for different resolutions. In this figure, the initial drop, which represents the exact solution, is also included for comparison. It is clear that as the resolution is increased, the simulation results converge towards the exact solution.

Following [13], we define the error of the simulation as

$$E_{shape} = \sum_{i=1}^{N_x} \sum_{j=1}^{N_y} |\phi(i, j, t = T) - \phi(i, j, t = 0)| \Delta x \Delta y. \quad (34)$$

Figure 11 shows that this error is reduced with approximately second order convergence rate. A comparison of these errors against the values reported by [13] is presented in Section 3.3.3.

*3.3.2. Zalesak disk*

This test case also measures the accuracy of advection techniques for two-phase flows. A notched disk with radius $R = 0.15$ and notch width and length of $w = 5$ and $l = 25$ is placed inside a $100 \times 100$ box with an imposed velocity field of $\vec{u} = (\pi(50 - y)/314, \pi(x - 50)/314)$. Theoretically, this velocity field, which corresponds to solid body rotation, should maintain the shape of the disk until it finally returns it to its initial condition at $t = 628$. A fixed time-step of $\Delta t = \Delta x/4$ is used for RK4 time integration. Figure 12 shows the outcome of simulations with different resolution of the Zalesak disk problem after the disk has rotated once. We refine the mesh while maintaining $\epsilon/\Delta x = 0.7$ and $\gamma/|\vec{u}|_{max} = 2.5$ fixed.

It is clear that as the mesh is refined, the simulation results become closer to the exact solution.

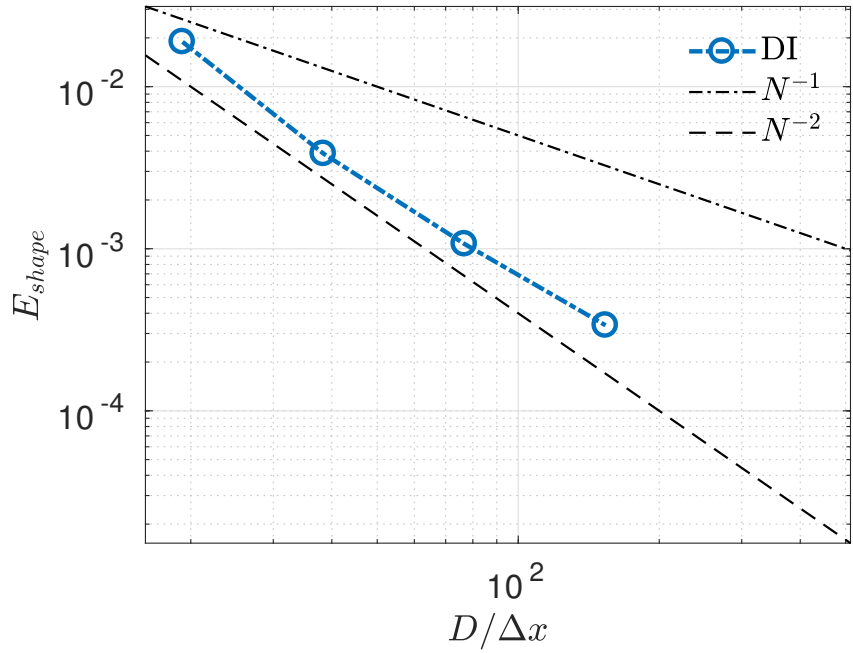

Figure 11: Shape error for the deforming drop test case versus resolution

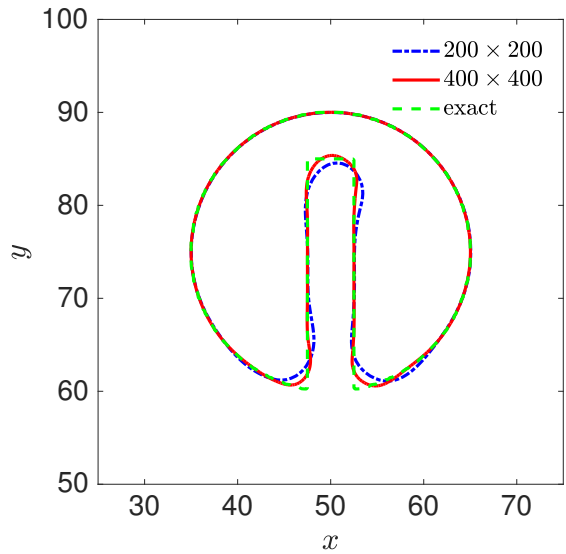

Figure 12: Zalesak disk after rotating once in simulations performed at different resolutions plotted against the exact solution.

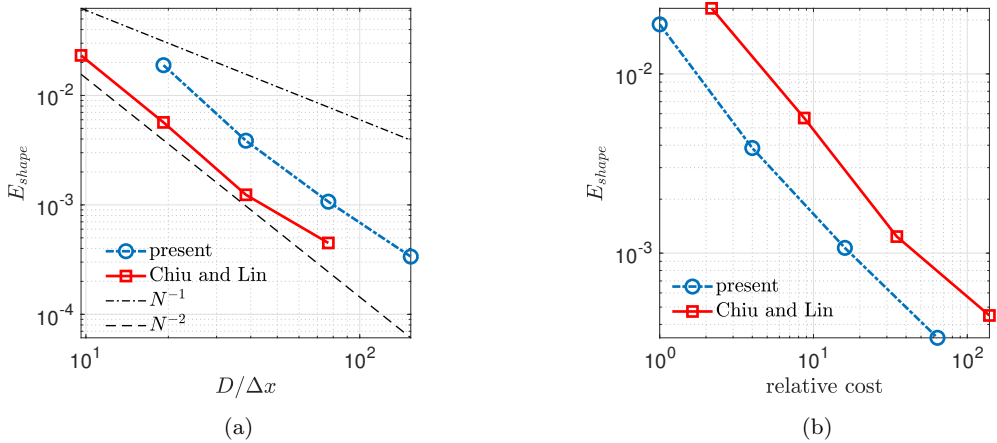

Figure 13: The accuracy of the present methodology compared to the method in [13] for different resolution simulations of the drop in shear flow test case, explained in Section 3.3.1 (a). Accuracy of the present methodology and the method of [13] plotted versus computational cost (b).

*3.3.3. Comparison of accuracy against the method of Chiu and Lin*

Chiu and Lin [13] reported accuracy versus resolution for the drop in shear flow test case. In our study, presented in Section 3.3.1, we adopted the same physical parameters, as well as the same measure to quantify the error (Equation 34). We note that [13] used a dual-compact dispersion preserving upwind advection scheme for computation of advective and diffusive fluxes with fifth and sixth order accuracy, respectively. This required solving linear systems that made their approach much more expensive than our proposed second order discretization scheme. Figure 13a compares error versus resolution between the present method and that of [13]. Despite the use of higher order fluxes by [13], it is interesting that our method preserves the same order of accuracy with only a factor of 2-3 penalty in the magnitude of errors.

More interestingly, when accuracy is plotted against computational cost, it is apparent from Figure 13b that our method yields more cost-effective solutions for the same accuracy demands. To estimate the relative cost, for each numerical scheme we count the number of floating point operations required for evaluation of $\partial \phi / \partial t$ in Equation 1. The proposed numerical scheme in this work does not involve any matrix inversions, and only requires explicit compact local stencils. Considering the operations needed for differentiation, computations of the normal vector, interpolation, multiplication and term addition we estimate the total number of operations to be 17 per mesh point per direction for each evaluation of $\partial \phi / \partial t$. On the other hand, the high order scheme of [13] requires solving two septa-diagonal linear systems for differentiation and interpolation routines for each evaluation of $\partial \phi / \partial t$ per direction. The number of unknowns in each of these linear systems is twice the number of mesh points per direction. Considering the steps required for forward elimination and back-substitution for direct solves, we estimate 68 floating point operations per mesh point for solving each $1D$ linear system. Given the required two linear systems per direction and the remaining operations for computing $\partial \phi / \partial t$, we estimate the total number of operations to be 148 per mesh point per direction. Based on this analysis, a factor of 148/17 is considered to convert the total number of mesh points to a measure of cost in Figure 13b.

It is important to remind that the described cost-vs-accuracy advantage is a bonus benefit of the present method. Namely, our method has crucial robustness properties, is simple to implement and develop upon, and is highly scalable on parallel platforms:

- The solution obtained from our method is guaranteed to be bounded as long as $\epsilon$ and $\gamma$ are chosen based on 17, while [13] were forced to clip out of bound values and redistribute the mass to conserve $\phi$.

- The explicit nature and its use of narrow numerical stencils makes the present method easy to implement and an attractive option compared to the much more involved schemes required by most accurate two-phase flow methods in literature.

- The present methodology can be fairly easily extended to non-uniform anisotropic Cartesian grids. A conservative strategy (but not necessarily an efficient one) would be to choose $\epsilon$ based on the worst case scenario, such that no interfacial region would go below the crossover line (Equation 17). A more optimal strategy is tuning $\epsilon$ based on the local mesh. Our investigations (out of the scope of this report) show that this retains the boundedness and accuracy properties of the method. On the other hand, the matrix coefficients presented in [13] are limited to uniform mesh spacings.

- The advection procedure outlined here is easily parallelizable and scalable. Conversely for the method of [13], parallel implementation of the high order differentiation and interpolation routines, which require solving septa-diagonal linear systems, demand significant transfer of data between cores, resulting in lower scalability.

## 4. Conclusions

This work proposes a conservative and bounded diffuse interface method for simulating incompressible two-phase flows which employs non-dissipative numerical discretizations. By examining discrete solutions to the diffuse interface evolution PDE (Eqaution 1), we identified boundedness conditions when a uniform staggered mesh is used,

$$\epsilon/\Delta x \geq \frac{\gamma/|\vec{u}|_{max} + 1}{2\gamma/|\vec{u}|_{max}}, \tag{35}$$

where the equality condition defines the crossover line in the $\epsilon/\Delta x - \gamma/|\vec{u}|_{max}$ parameter space. While the detailed proof was presented for explicit Euler time-stepping in multi-dimensions, we explained and confirmed how this condition is more broadly applicable to higher order Runge-Kutta time-advancement schemes.

Throughout this work we also provided guidelines for users interested in adopting the phase field method introduced herein. Using linear stability analysis, we demonstrated that the crossover line corresponds to the ridge of maximum stable time-steps. Thus for bounded simulations $\epsilon/\Delta x$ and $\gamma/|\vec{u}|_{max}$ should be chosen from points on the crossover line. On the crossover line, there is a trade-off between numerical accuracy (lowering $\epsilon$ reduces thickness of interface) and temporal stiffness (higher $\epsilon$ values allow for smaller CFL) which guides the user in choosing an appropriate $\epsilon/\Delta x - \gamma/|\vec{u}|_{max}$ pair.

Additionally, using a simple 1D advection test case we presented a general procedure for finding optimal convergence strategy for diffuse interface methods. While for the simple passive problem studied here, it was recommended to fix $\epsilon/\Delta x$ and $\gamma/|\vec{u}|_{max}$ while refining the mesh, the same procedure can potentially lead to other recommendations for other problems, especially when Equation 1 is coupled to the Navier-Stokes equation.

Furthermore, we provided two-dimensional numerical tests to firstly verify that our analytical proof holds in extreme numerical settings, and secondly to demonstrate the accuracy of the presented discretization of Equation 1. A comparison of accuracy against the method of [13] revealed that when accuracy is assessed against numerical cost, the present method offers a significant improvement while maintaining the same overall order of accuracy.

Because the phase field representation of the interface is inherently smooth, the spatial discretization utilizes central differences, and the temporal discretization is fully explicit, this approach is naturally load balanced and scalable. As such, this simple, cheap, parallelizable, conservative, bounded and accurate interface capturing scheme is presented as a viable approach for simulating realistic two-phase flows.

## 5. Acknowledgements

This work was supported by Office of Naval Research (Grant No. 119675) and NASA (Grant No. 127881) .